\documentclass[%
 aip,
 amsmath,amssymb,
reprint,%
]{revtex4-1}

\usepackage{graphicx}
\usepackage{dcolumn}
\usepackage{bm}

\usepackage[colorlinks=true,
            linkcolor=blue,
            urlcolor=blue,
            citecolor=blue]{hyperref}
\usepackage{svg}
\usepackage{caption}
\usepackage{floatrow}

\usepackage[utf8]{inputenc}
\usepackage[T1]{fontenc}
\usepackage{mathptmx}
\usepackage{etoolbox}

\def\fullnameCap{Parameter inference from a non-stationary unknown process}
\def\fullnameAllCap{Parameter Inference from a Non-stationary Unknown Process}

\def\problemname{PINUP}

\makeatletter
\def\@email#1#2{%
 \endgroup
 \patchcmd{\titleblock@produce}
  {\frontmatter@RRAPformat}
  {\frontmatter@RRAPformat{\produce@RRAP{*#1\href{mailto:#2}{#2}}}\frontmatter@RRAPformat}
  {}{}
}%
\makeatother
\begin{document}

\preprint{AIP/123-QED}

\title[\fullnameCap{}]{\fullnameCap{}}

\author{Kieran S. Owens}
\author{Ben D. Fulcher}%
 \email{ben.fulcher@sydney.edu.au}
\affiliation{%
 School of Physics, The University of Sydney, Camperdown NSW 2006, Australia
}%
\affiliation{Centre for Complex Systems, 
 The University of Sydney, Camperdown NSW 2006, Australia.}

\date{\today}

\begin{abstract}
Non-stationary systems are found throughout the world, from climate patterns under the influence of variation in carbon dioxide concentration, to brain dynamics driven by ascending neuromodulation.
Accordingly, there is a need for methods to analyze non-stationary processes, and yet most time-series analysis methods that are used in practice, on important problems across science and industry, make the simplifying assumption of stationarity.
One important problem in the analysis of non-stationary systems is the problem class that we refer to as \fullnameAllCap{} (\problemname{}).
Given an observed time series, this involves inferring the parameters that drive non-stationarity of the time series, without requiring knowledge or inference of a mathematical model of the underlying system.
Here we review and unify a diverse literature of algorithms for \problemname{}.
We formulate the problem, and categorize the various algorithmic contributions into those based on: (1) dimension reduction; (2) statistical time-series features; (3) recurrence quantification analysis; (4) prediction error; (5) phase-space partitioning; and (6) Bayesian inference.
This synthesis will allow researches to identify gaps in the literature and will enable systematic comparisons of different methods.
We also demonstrate that the most common systems that existing methods are tested on---notably the non-stationary Lorenz process and logistic map---are surprisingly easy to perform well on using simple statistical features like windowed mean and variance, undermining the practice of using good performance on these systems as evidence of algorithmic performance.
We then identify more challenging problems that many existing methods perform poorly on and which can be used to drive methodological advances in the field.
Our results unify disjoint scientific contributions to analyzing non-stationary systems and suggest new directions for progress on the \problemname{} problem and the broader study of non-stationary phenomena.
\end{abstract}

\maketitle

\begin{quotation}
The analysis of time-series data is important for a range of fields, including physics, neuroscience, ecology, and finance.
Time-series analysis methods commonly assume that the measured process is stationary, i.e., that certain properties of the process, such as mean and variance, are constant across time.
And yet many important processes are non-stationary on relevant dynamical timescales.
For example, time series measured from electrical activity of the human brain during sleep exhibit considerable variation in statistical properties across wakefulness and different stages of sleep.
Thus there is a need for methods that can quantify non-stationary dynamical variation.
Here we consider the specific problem of inferring the parameters that drive non-stationarity of a time series, without requiring knowledge or inference of a mathematical model of the underlying system.
We call this problem \fullnameAllCap{} (\problemname{}).
We make two main contributions.
First, we review and categorize the algorithmic approaches to \problemname{} across a fragmented literature, consolidating a diverse body of methodological research for the first time.
Unifying the terminology of the problem and providing an overview of the scientific methods that have been developed to date will allow researchers to take a unified approach to targeting gaps in the interdisciplinary literature, thus facilitating progress on \problemname{}.
Second, we show that the systems most commonly used for benchmarking---the Lorenz process and logistic map---admit trivial algorithmic solutions, undermining the practice of presenting good performance on these problems as evidence of algorithmic progress.
We present more challenging test problems, which many existing methods perform poorly on, and that can be used to drive improvements in \problemname{} algorithms. 
In turn, this will foster the development of new approaches to the broader study of non-stationary phenomena.
\end{quotation}

\section{Introduction}
\label{sec:level1}

Time-series data are ubiquitous across science and industry, making the tools of time-series analysis essential for contemporary analytic practice.
To make analyses tractable, time series are commonly assumed to be stationary.
The assumption of \textit{weak stationarity} entails that the first and second moments of the distribution of a time series are constant, whereas \textit{strong stationarity} means that all conditional probabilities are constant in time. \cite{schreiberClassificationTimeSeries1997}
However, many studied systems exhibit non-stationary behavior that violates one or both of these assumptions.
Representative examples of non-stationary systems can be found in physics, \cite{bucaNonstationaryCoherentQuantum2019} engineering, \cite{chelidzeFatigueDamage2005} ecology, \cite{summersChaosPeriodicallyForced2000} climate science, \cite{zhangReconstruction2017} neuroscience, \cite{galadiNonstationaryBrain2021} 
and finance.\cite{schmittNonstationarityFinancialTime2013}
Moreover, for many studied systems the \textit{pattern} of non-stationary variation in dynamical behavior is of interest.
For example, in the neuroscientific context it is useful to ask whether instances of non-stationary brain dynamics are driven by patterns of activity in certain brain regions. \cite{shineNeuromodulatoryControlComplex2023a}
The non-stationarity of such systems can be conceptualized in terms of one or more processes that vary the conditional probabilities of the system by modulating parameters of a hypothetical model or set of generative equations.
Further, if we can reconstruct a time series of the parameter variation underlying non-stationarity, we can then seek a correspondence between this parameter time series and some part of the system or its environment.
In some cases, assuming stationarity may not only be incorrect, but can blind us to important sources of variation that are crucial to understanding the system.
For example, a recent study suggests that the time-series features that best discriminate whether a subject is an experienced meditator using electroencephalography (EEG) are those that index non-stationarity, with traditional spectral properties showing no difference. \cite{baileyUncoveringStabilitySignature2024}
In summary, there is a need to develop methods that can be used to analyze non-stationary systems---e.g., by identifying and quantifying non-stationarity, and characterizing the dynamics of processes that drive the statistical variation.

A pragmatic definition of stationarity is needed for data-driven analyses because, as theoretically defined for stochastic processes, stationarity is a property of processes rather than of finite time series, and can only be evaluated in the large data limit---in the univariate case, $x_t$ ($t = 1, \cdots, T$), as the number of samples $T \rightarrow \infty$.
When stationarity is discussed in relation to deterministic dynamical systems, an appeal is often made to the concept of ergodicity, which is also problematic for finite data.\cite{ruelleChaoticEvolutionStrange1989}
Additionally, there is no consensus theoretical definition of stationarity for deterministic dynamical systems: some authors require the existence of an invariant ergodic measure,\cite{kennelStatisticalTestDynamical1997,
wittTestingStationarityTime1998,
heggerCopingNonstationarityOverembedding2000} some require that the system be autonomous (i.e., with equations and parameters that are fixed across time),\cite{manucaStationarityNonstationarityTime1996, riekeMeasuringNonstationarityAnalyzing2002,
serquinaCharacterizationNonstationaryChaotic2008,
bolltControlEntropyComplexity2009,
rayCharacterizationNonstationaryChaotic2010} while others require that time-series statistics do not vary over the duration of measurement.\cite{schreiberDetectingAnalysingNonstationarity1997, aguirreNonstationaritySignaturesDynamics2012a}
Similar to \citet{schreiberDetectingAnalysingNonstationarity1997}, and \citet{aguirreNonstationaritySignaturesDynamics2012a}, here we adopt a definition of non-stationarity as the variation over time in the joint probability distribution of a time series, for which variation in dynamical properties may be used as a surrogate measure.
In practice, measuring variation in joint probabilities or certain statistics necessitates specifying some \textit{timescale}, i.e., some temporal window length over which this variation is evaluated, and in general stationarity is \textit{timescale dependent}.
For example, a process with a constant mean over long time windows can have a wandering mean over short windows.
What timescale to consider is best decided by a practitioner on the basis of domain knowledge.
For example, given an EEG time series measured during sleep, the relevant timescale will be much shorter if 200\,Hz hippocampal ripples \cite{siapasCoordinatedInteractionsHippocampal1998} are the object of study, as compared to sleep stages, which by convention are rated using 30\,s epochs \cite{moserSleepClassificationAccording2009}.

In this work we consider the problem of \fullnameAllCap{} (\problemname{}), for which we adopt the following definition (which we develop more formally in Sec.~\ref{sec:problemFormulation}).
Starting with time-series observations of a non-stationary process, and in the absence of a known mathematical model of the process, \problemname{} involves inferring one or more time-varying parameters (TVPs) that drive the non-stationarity. 
A \problemname{} method is any algorithmic procedure for solving this problem.
The specification that the generative process be unknown \textit{a priori} is important.
Firstly, there are many real-world settings where there is considerable uncertainty about the functional form of a process.
For example, for complex systems such as the brain, where we often have large amounts of noisy, time-series data, finding relevant time-varying patterns within the data is often more feasible than developing an accurate model of the entire system.
Secondly, it is important to distinguish \problemname{} from the problem of inferring static or time-varying parameters for a \textit{known} model, an example of which is the inverse problem for differential equations.\cite{tarantolaInverseProblemTheory2005}
The Kalman filter, particle filter, and other Bayesian and statistical inferential methods have been developed for this purpose.\cite{sarkkaBayesianFilteringSmoothing2023,
gunturkunRecursiveHiddenInput2014,
arnoldApproachPeriodicTimevarying2018,
hauzenbergerFastFlexibleBayesian2022,
arnoldWhenArtificialParameter2023}
In some cases it may be possible to reduce \problemname{} to this latter problem class if a generative model can first be learned via data-driven discovery.\cite{luchinskyInferentialFrameworkNonstationary2008,
bruntonDiscoveringGoverningEquations2016a,
northReviewDataDrivenDiscovery2023, 
nicolaouDatadrivenDiscoveryExtrapolation2023b}

\begin{figure*}[hbt!]
\includegraphics[width=1.0\textwidth]{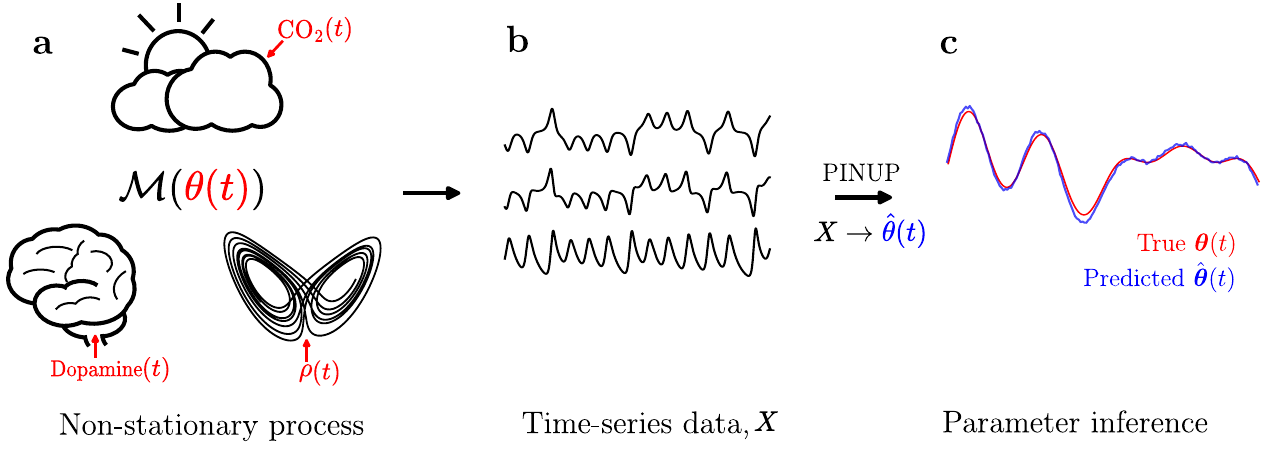}

\caption{\label{fig:Summary}
\textbf{\fullnameAllCap{} (\problemname{})}. 
\textbf{a.} A non-stationary process is generated by an unknown model $\mathcal{M}$ under the influence of some time-varying parameter(s) (TVPs) $\boldsymbol{\theta}(t)$. 
A selection of example processes are depicted with TVPs denoted in red:
(i) brain activity under the time-varying influence of ascending neuromodulation by dopamine;
(ii) climate (e.g., mean temperature) under the influence of changing $\text{CO}_2$; and
(iii) the Lorenz process with a time-varying parameter, $\rho(t)$ [see Eq.~\eqref{eq:lorenz}].
\textbf{b.} A time-series realization $X$ of the non-stationary process is observed.
\textbf{c.}
A \problemname{} algorithm is used to infer the time-varying parameter(s) $\hat{\boldsymbol{\theta}}(t)$ directly from the time series $X$.
}
\end{figure*}

\problemname{} is an important problem that has been studied widely across different disciplinary contexts, resulting in a correspondingly fragmented literature on approaches to tackling it.
Existing theory and algorithms that have been developed for it are broad, from dimension reduction to Bayesian inference.
These myriad theoretical and algorithmic contributions are often disjoint, with overall minimal referencing of approaches developed in different fields, suggesting a lack of familiarity with the interdisciplinary literature.
There is no agreed problem formulation, and variation in terminology---e.g., inferring a ``drifting parameter", \cite{chatterjeeOptimalTrackingParameter2002} or a ``driving force", \cite{verdesNonstationaryTimeSeriesAnalysis2001} versus a ``causal driver" \cite{gilpinRecurrencesRevealShared2023}---further compounds this.
Furthermore, many papers only demonstrate the usefulness of their individual algorithm on a single problem, most commonly the logistic map or the Lorenz process.\cite{casdagliRecurrencePlotsRevisited1997,
      verdesNonstationaryTimeSeriesAnalysis2001,
      chatterjeeOptimalTrackingParameter2002,
      szeligaExtractingDrivingSignals2002,
      szeligaModellingNonstationaryDynamics2003,
      wiskottEstimatingDrivingForces2003,
      verdesOverembeddingMethodModeling2006,
      gunturkunSequentialReconstructionDrivingforces2010,
      slobodaSensitivityVectorFields2013,
      grinblatNonstationaryRegressionSupport2015,
      niknazarVolumetricBehaviorQuantification2017a,
      wangRobustnessDrivingForce2023} 
In addition to the challenge of determining the current state of the interdisciplinary scientific literature on this topic, this lack of benchmarking standards or systematic comparison of algorithm performance across a diverse range of challenging systems (as is common practice for other problem classes) makes it difficult for a reader to discern whether progress is being made.
Here we unify the disjoint literature on \problemname{} by clearly defining the problem class (Sec.~\ref{sec:problemFormulation}) and reviewing and categorizing the diverse, algorithmic, scientific literatures (Sec.~\ref{sec:existingMethods}).
We then demonstrate, using new numerical experiments, the trivial nature of problems commonly used for benchmarking \problemname{} algorithms---including the logistic map---and introduce more challenging problems on which \problemname{} performance may more appropriately be tested, and on which many conventional methods fail (in Sec.~\ref{experiment}).

\section{Problem formulation}
\label{sec:problemFormulation}

The basic procedure of \problemname{} is illustrated in Fig.~\ref{fig:Summary}.
\problemname{} aims to infer the time series of one or more sources of non-stationary variation directly from a measured (univariate or multivariate) time series.
Taking an example from Fig.~\ref{fig:Summary}, suppose that the dynamics of a brain were to be deliberately manipulated by modulating brain dopamine activity over time.
Given a time series of brain activity, such as EEG data, one could use \problemname{} to try to infer a time series of dopamine activity.
Similarly, if a chaotic flow such as the Lorenz process has a time-varying parameter (TVP), then we can analyze the Lorenz process time series to try to infer the time series of the TVP.

We consider a time series $\boldsymbol{x}(t)$ of a non-stationary process to be generated by a model $\mathcal{M}$, the joint probability distribution of which depends on one or more time-varying parameters, $\boldsymbol{\theta}(t)$, across some time interval $t \in [0,T_{\text{max}}]$.
By framing \problemname{} in terms of an general model $\mathcal{M}$, we allow for a range of generative processes.
For example, $\mathcal{M}$ could represent a system of ODEs $\frac{d}{dt} \boldsymbol{x} = f(\boldsymbol{x}, \boldsymbol{\theta}(t))$, or an iterative map $\boldsymbol{x}_{t+1} = f(\boldsymbol{x}_t, \boldsymbol{\theta}_t)$, among other possibilities including noise processes, stochastic differential equations (SDEs), or even a process governed by rules that could not be written down in closed form.
Stationarity then corresponds to the fixed-parameter case, $\boldsymbol{\theta}(t) = \boldsymbol{\theta}(0)$, $\forall t$.
In practice, regardless of whether $\mathcal{M}$ is defined in continuous or discrete time, to obtain a time series of length $T$, the process must be sampled discretely, for example, at a uniform sampling rate as $t = 0, \Delta t, 2\Delta t, ..., (T-1) \Delta t$.
We denote a time-series realization from $\mathcal{M}(\boldsymbol{\theta}(t))$ as $X \in \mathbb{R}^{T \times D}$, where $D$ is the number of time-series variables, and where the value at time $t$ of variable $i$ is $x_{t, i}$. 
Given an ascending sequence of sampling times $t_i \in [0,T_{\text{max}}]$, we let $\boldsymbol{x}_i \equiv \boldsymbol{x}(t_i)$ and can write $X$ in terms of row vectors:
\begin{equation} X = 
\begin{bmatrix}
    \rule[.5ex]{1em}{0.4pt} & \boldsymbol{x}_1 & \rule[.5ex]{1em}{0.4pt}\\
    \rule[.5ex]{1em}{0.4pt} & \boldsymbol{x}_2 & \rule[.5ex]{1em}{0.4pt}\\
    & \vdots \\
    \rule[.5ex]{1em}{0.4pt} & \boldsymbol{x}_T & \rule[.5ex]{1em}{0.4pt}
\end{bmatrix}
\end{equation}
The corresponding discretization in time of $\boldsymbol{\theta}(t)$ is $\boldsymbol{\theta}_t$.
More generally, if there is some observation function $g$ and noise process $\boldsymbol{\eta}$, then our observed time series is $Y \in \mathbb{R}^{T \times D}$, where $\boldsymbol{y}_{t} = g(\boldsymbol{x}_{t}) + \boldsymbol{\eta}_{t}$. 
We obtain $Y = X$ if observations are noise-free and $g$ is the identity function.

\problemname{} can be summarized as follows: given an observed time-series dataset $X$ (or $Y$), resulting from some \textit{unknown} process $\mathcal{M}$ under the influence of TVP(s) $\boldsymbol{\theta}(t)$, our goal is to infer an approximation of the TVP(s), denoted $\hat{ \boldsymbol{\theta}}_t$.

Although many of the papers in the \problemname{} literature refer to the inference of a time-varying parameter, other terminology includes inference of a ``driving force", \cite{szeligaModellingNonstationaryDynamics2003, wiskottEstimatingDrivingForces2003, tanioReconstructionDrivingForces2009, gunturkunSequentialReconstructionDrivingforces2010} a ``driving signal", \cite{szeligaExtractingDrivingSignals2002} a ``drifting parameter", \cite{chatterjeeOptimalTrackingParameter2002, chelidzeDynamicalSystemsApproach2002, grinblatNonstationaryRegressionSupport2015} a ``dynamical parameter", \cite{alaoDiscoveringDynamicalParametersa} or a ``causal driver". \cite{gilpinRecurrencesRevealShared2023}
In what follows we treat these terms as synonyms and use the term TVP for consistency.

\section{Review of existing methods}
\label{sec:existingMethods}

How might one approach \problemname{}?
Starting from the concept of non-stationarity, one could proceed by developing a method that quantifies variation in the joint probability distribution and then uses this information to infer one or more TVPs.
For example, some statistics (or `features') of the dynamics will be sensitive to changes in the joint probability distribution induced by $\boldsymbol{\theta}_t$ (and in the optimal case will be sufficient statistics with respect to $\boldsymbol{\theta}_t$).
Therefore, one could measure a set of time-series statistics across sliding time-series windows and then examine the resulting time series of statistics to obtain an estimate of the TVP(s). \cite{guttlerReconstructionParameterSpaces2001}
Similarly, one could compute windowed empirical probability distributions, say, through naive histogram binning, in order to estimate variation in the underlying probability density function (for a stochastic process) or invariant measure (for a deterministic process). \cite{carrollAttractorComparisonsBased2015}
Moreover, if a pattern of non-stationary variation manifests jointly across several time-series variables (or statistical features) then dimension reduction could be used to project the data onto a lower-dimensional space that captures shared variation that is driven by the underlying parameter(s). \cite{wiskottEstimatingDrivingForces2003}
Indeed, many such \problemname{} algorithms have been proposed but they have never been systematically studied or compared.

In this work we organize the literature of theory and algorithms for \problemname{} into six categories of conceptually distinct approaches, illustrated in Fig.~\ref{fig:Methods}, which we explain in turn through this section.
In particular, we organize existing methods into those based on:


\begin{figure*}
\ffigbox{}{\CommonHeightRow{\begin{subfloatrow}[3]
\captionsetup{justification=centering}
\ffigbox[\FBwidth]
{\includegraphics[height=\CommonHeight]{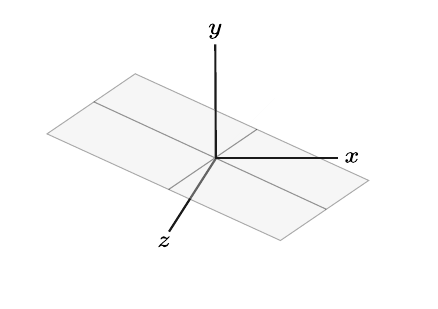}}{\caption*{a. Dimension reduction}}
\ffigbox[\FBwidth]
{\includegraphics[height=\CommonHeight]{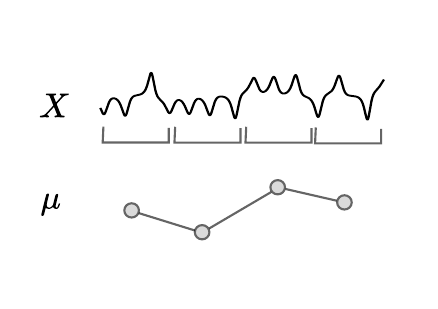}}{\caption*{b. Statistical time-series features}}
\ffigbox[\FBwidth]
{\includegraphics[height=\CommonHeight]{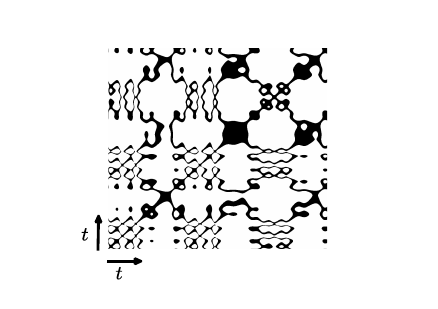}}{\caption*{c. Recurrence quantification analysis}}
\end{subfloatrow}
\begin{subfloatrow}[3]
\captionsetup{justification=centering}
\ffigbox[\FBwidth]
{\includegraphics[height=\CommonHeight]{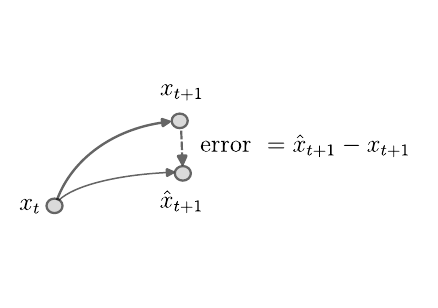}}{\caption*{d. Prediction error}}
\ffigbox[\FBwidth]
{\includegraphics[height=\CommonHeight]{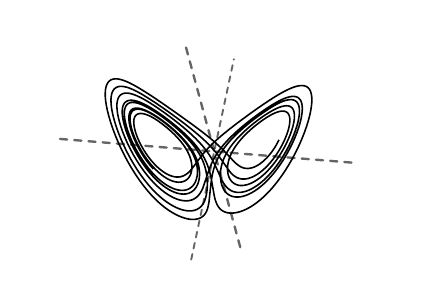}}{\caption*{e. Phase-space partitioning}}
\ffigbox[\FBwidth]
{\includegraphics[height=\CommonHeight]{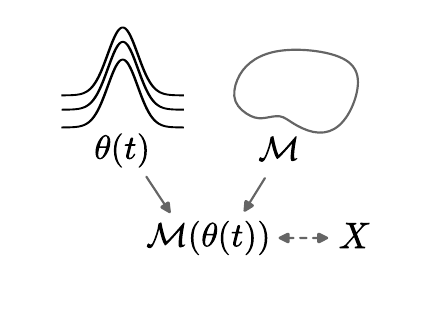}}{\caption*{f. Bayesian inference}}
\end{subfloatrow}}
\caption{\label{fig:Methods}\textbf{Schematic depiction of six categories of \problemname{} methods, each of which tracks non-stationarity through some aspect of parameter-driven variation in the joint probability distribution of a process.}
\textbf{a.} \textit{Dimension reduction} methods project time-series data onto lower-dimensional spaces that optimize with respect to some statistical property (such as variance, slowness, autocorrelation, or predictability). 
The image shows data being mapped to a $2$-dimensional subspace via a linear transformation.
\textbf{b.} \textit{Statistical time-series feature} methods infer parameter variation from statistical variation. The figure depicts mean values of a single time-series variable from the Lorenz process computed using sliding windows.
\textbf{c.} \textit{Recurrence quantification analysis} methods infer parameter variation from recurrence plots which are computed based on the patterns of recurrence in time-series data. The recurrence plot of a Lorenz process driven by a sinusoidal parameter $\rho$ is shown.
\textbf{d.} \textit{Prediction error} methods construct a time-series prediction/forecasting model and then infer parameter variation from the resulting prediction error time series. The figure depicts one-step prediction error between the true trajectory and a prediction using some function.
\textbf{e.} \textit{Phase space partitioning} methods infer parameter variation from multivariate time series of metrics, such as prediction error, that are measured across different regions. The figure depicts a partition of the phase space of a Lorenz process.
\textbf{f.} \textit{Bayesian inference} methods start from some prior distribution over models and TVP values and then infer parameter variation from the data on the basis of a likelihood model.
In the figure, parameter values $\theta (t)$ and a model $\mathcal{M}$ are drawn from prior distributions, then data realizations generated from a process $\mathcal{M}(\theta(t))$ are compared to the observed time series $X$.}}
\end{figure*}%



\begin{itemize}
    \item \textit{Dimension reduction} (Sec.~\ref{sec:dim_red}, Fig.~\ref{fig:Methods}a).
    These methods project time-series data onto lower-dimensional spaces that best capture some property (such as variance, autocorrelation, or predictability) that is sensitive to underlying TVPs.
    
    \item \textit{Statistical time-series features} (Sec.~\ref{sec:stat_ts_features}, Fig.~\ref{fig:Methods}b). These methods track parameter variation using statistics that may be sensitive to variation in the joint probability distribution of the underlying process.

    \item \textit{Recurrence quantification analysis} (RQA) (Sec.~\ref{sec:rqa}, Fig.~\ref{fig:Methods}c).
    These methods infer TVPs from recurrence plots which are computed based on the patterns of recurrence in time-series data.
    
    \item \textit{Prediction error} (Sec.~\ref{sec:pred_error}, Fig.~\ref{fig:Methods}d).
    These methods construct a time-series prediction/forecasting model and then infer TVPs from the time series of prediction error.
    
    \item \textit{Phase-space partitioning} (Sec.~\ref{sec:phase_space}, Fig.~\ref{fig:Methods}e).
    These methods infer TVPs from multivariate time series of metrics, such as prediction error, that are measured across different regions of abstract space.
    
    \item \textit{Bayesian inference} (Sec.~\ref{sec:bayes}, Fig.~\ref{fig:Methods}f).
    These methods start from some prior distribution and then infer TVPs from the data using a probability model.
    
\end{itemize}

\subsection{Dimension reduction}
\label{sec:dim_red}

Dimension reduction involves mapping data to a lower-dimensional space in a way that optimizes or preserves some property of the data. \cite{burgesDimensionReductionGuided2009}
For example, principal component analysis (PCA), learns linear projections that maximize variance or, equivalently, minimize squared reconstruction error. \cite{jolliffeianPrincipalComponentAnalysis2002, udellGeneralizedLowRank2016}
The structure of time-series data, namely, that the data are ordered and will in general display autocorrelation, poses both challenges and opportunities for dimension reduction methods.
For example, time-series autocorrelation violates the assumption of independent and identically distributed (i.i.d.) data used by many methods, but can also form the basis of alternative methods that exploit this structure (as seen below).
For \problemname{}, if non-stationary variation manifests across a multivariate time series $X$---say, by modulating variance or autocorrelation---then a dimension-reduction method that is sensitive to such changes could be used to project onto one or more dimensions that track the TVPs.
This idea is depicted schematically in Fig.~\ref{fig:Methods}a, which depicts a linear dimension reduction from three dimensions to a two-dimensional subspace.

Dimension reduction was used by \citet{wiskottEstimatingDrivingForces2003} to infer parameter variation from non-stationary time series.
To do this, Wiskott used Slow Feature Analysis (SFA), which extracts slowly varying features for which the first temporal derivative is minimized. \cite{wiskottSlowFeatureAnalysis2002}
In contrast to PCA, the first dimension of which identifies the direction of greatest time-series variance, SFA tracks the direction of \textit{slowest} variation in the time series.
The inductive bias of SFA towards slowness makes it sensitive to non-stationary dynamical variation when fast observed dynamics are modulated by a slower TVP.
When SFA is used for \problemname{}, time-delay embedding \cite{kantzNonlinearTimeSeries2003} (discussed further in Sec.~\ref{sec:pred_error}) and polynomial basis expansion are first applied \cite{wiskottSlowFeatureAnalysis2002} (noting that the method is termed `SFA2' when a second-order polynomial basis expansion is used).
The noise-robustness of SFA for parameter inference was recently examined by \citet{wangRobustnessDrivingForce2023a}, and numerous SFA variants have been applied to a range of problems over the past two decades. \cite{songSlowGoBetter2022}
A closely related dimension reduction method is Time-lagged Independent Component Analysis (TICA) which finds components that have the largest autocorrelation at a given time-lag. \cite{molgedeySeparationMixtureIndependent1994a,
      schultzeTimeLaggedIndependentComponent2021a}
TICA is equivalent to SFA, without TDE or basis expansion, in the case where one-step autocorrelation is considered. \cite{blaschkeWhatRelationSlow2006a}
Several papers have explored the mathematical relationships between techniques such as PCA, SFA, TICA, Smooth Orthogonal Decomposition (SOD, detailed in Sec.~\ref{sec:phase_space}), and other methods.
\cite{blaschkeWhatRelationSlow2006a,
      klusDataDrivenModelReduction2018a,
      khanUnifiedInterpretationProper2020a}
Nonlinear neural-network extensions of these and similar dimension-reduction techniques have also been proposed, such as deep-TICA, Time-lagged Auto-Encoders (TAEs), and Variational Approach for Markov Process networks (VAMPnets), many of which have been applied to discover slow collective variables underlying molecular dynamics.
\cite{mardtVAMPnetsDeepLearning2018,
      wehmeyerTimelaggedAutoencodersDeep2018,
      bonatiDeepLearningSlow2021a} 
In contrast to SFA2, which applies a polynomial basis expansion by default,
\cite{wiskottSlowFeatureAnalysis2002} deep-TICA \textit{learns} and applies a nonlinear basis expansion followed by TICA decomposition using a differentiable network layer. \cite{bonatiDeepLearningSlow2021a} 
The linear methods, such as PCA, SFA, and TICA, trade expressivity for interpretability, noting that linear transformation weights can be interpreted as indicating the relative importance of different variables or features for each dimensional projection.
In contrast, the nonlinear methods can flexibly learn more complex mappings but pose a risk of overfitting, especially for small datasets or when the signal-to-noise ratio is low.
\cite{hastieElementsStatisticalLearning2017}

Several dimension reduction \problemname{} methods have been applied to non-stationary time series to infer TVPs as a part of scientific discovery. 
For example, SFA has been used in climate science to infer TVPs that drive climate change and other variation related to greenhouse and aerosol emissions,\cite{verdesGlobalWarmingDriven2007}
the North Atlantic Oscillation,\cite{wangExtractingDrivingForce2016}
the El Niño-Southern Oscillation and Hale sunspot cycles,\cite{wangIdentificationDrivingForces2017} 
the Atlantic Multi-decade Oscillation,\cite{yangCausalityGlobalWarming2016,
zhangReconstructionDrivingForces2017} 
and the East Asian trough.\cite{luCaseStudiesDriving2020}
Additionally, there is an active literature using methods such as TICA, TAEs, and VAMPnets to discover reduced-order models of the patterns of variation underlying molecular activity.
\cite{naritomiSlowDynamicsProtein2011,
      perez-hernandezIdentificationSlowMolecular2013a,
      mardtVAMPnetsDeepLearning2018,
      paulIdentificationKineticOrder2019a,
      bonatiDeepLearningSlow2021a}

\subsection{Statistical time-series features}
\label{sec:stat_ts_features}

A statistical time-series feature is the scalar output of a function that has been applied to a time series, mapping a time series $\boldsymbol{x} \in \mathbb{R}^T \rightarrow \mathbb{R}$. \cite{fulcherFeatureBasedTimeSeriesAnalysis2018}
Thousands of such time-series features have been developed across a wide range of scientific disciplines, with simple examples including mean and variance, and more complex examples including entropy measures and nonlinear methods such as correlation dimension.
\cite{fulcherHctsaComputationalFramework2017a}
A feature vector can be computed by applying multiple algorithms to a time series, thereby providing a multifaceted statistical description of the time series.
Applying such algorithms across sliding time-series windows results in a time series of statistical features, which describes the variation in statistical properties over time.
Provided the chosen statistics are sensitive to TVP-driven statistical variation, time-series features can be used individually or in aggregate to infer TVPs.
When multiple time-series features are used, a natural step is to apply some form of dimension reduction in order to obtain the TVP estimate $\hat{\boldsymbol{\theta}}_t$. 
As an example, consider a hypothetical process in which a single TVP $\theta(t)$ varies the mean offset of the process. 
Simply computing a moving average will track $\theta(t)$ accurately.
Alternatively, if $\theta(t)$ were to modulate the frequency of an oscillatory component of a time series, then $\theta(t)$ could be tracked by computing variation in the peak location of the power spectrum over time.
Crucially, such windowed, feature-based inference of TVPs assumes that the dynamics within each window are pseudostationary, so that within-window parameter variation can be neglected.
This approach is illustrated in Fig.~\ref{fig:Methods}b as the computation of mean time-series values across windows.

Statistical features of a time series---such as mean, variance, and
autocorrelation---computed over sliding time-series windows were used by \citet{guttlerReconstructionParameterSpaces2001} to successfully reconstruct the space of TVPs of dynamical systems.
However, their method requires manual selection of features that are sensitive to the dynamical changes induced by the TVPs relevant to any given problem; it does not provide a feature set that is sensitive to parameter variation in general.
In a similar vein, other authors have performed \problemname{} using one or more hand-selected features.
For example, \citet{yansonGlobalReconstructionNonstationary1999} showed that a slow TVP driving a R\"{o}ssler attractor 
can be reconstructed using variance, and \citet{bolltControlEntropyComplexity2009} proposed a complexity measure termed `control entropy' with the aim of tracking TVPs.
Addressing the related problem of parameter variation \textit{across} a set of time series (each generated with some fixed set of parameter values) rather than over time, \citet{niknazarVolumetricBehaviorQuantification2017a} assessed the performance of individual nonlinear time-series features and volumetric features, which measure occupied space and geometric trajectory properties, for inferring parameter settings of the Lorenz 
and Mackey--Glass systems.
Inferring parameter variation across time series was similarly undertaken by \citet{alaoDiscoveringDynamicalParametersa}, who applied PCA dimension reduction to feature vectors of regression coefficients from the output layer of an echo state network, and by \citet{fulcherInferringLowdimensionalParametrica}, who addressed the issue of manual feature selection by applying PCA to a comprehensive set of over 7000 time-series features from the \textit{hctsa} package \cite{fulcherHctsaComputationalFramework2017a}.
Harris and Fulcher
\cite{harrisInferringParametricVariation2021} demonstrated that feature-based inference of a TVP can be biased by feature--feature redundancy, but that this issue can be ameliorated by applying PCA to appropriately chosen baseline data to find suitable orthogonal coordinates.

\subsection{Recurrence quantification analysis}
\label{sec:rqa}

Introduced by \citet{eckmannRecurrencePlotsDynamical1987}, recurrence plots (RP) represent the times at which states in a phase space recur over the course of a time series. 
Given a time series $X$, a threshold distance $\epsilon$, a norm $\lVert \cdot \rVert$, and the Heaviside function $\Theta(\cdot)$, a RP is constructed as a two-dimensional matrix $R$ according to $R_{i,j} = \Theta(\epsilon - \lVert \boldsymbol{x}_{i} - \boldsymbol{x}_{j} \rVert)$.
\cite{webberRecurrenceQuantificationAnalysis2015} 
Accordingly, a value of $1$ at $R_{i,j}$ means that the time-series states at times $i$ and $j$ are within distance $\epsilon$.
An example RP matrix for a non-stationary Lorenz process driven by a sinusoidal $\rho$ parameter is visualized in Fig.~\ref{fig:Methods}c.
Note that some form of phase-space reconstruction, typically time-delay embedding, is performed prior to constructing the RP, requiring the selection of additional hyperparameters. \cite{webberRecurrenceQuantificationAnalysis2015}
In this section we focus on methods that analyze the RP directly as a matrix or graph, whereas analyzing time series of RQA statistics would form a subset of the times-series feature-based methods grouped together in Sec.~\ref{sec:stat_ts_features}.
When applied to \problemname{}, the intuition is that the RP structure is an informative signature of the (possibly nonlinear) statistical properties of a process, and so non-stationarity should be reflected in the RP.
For example, if similar values of $\boldsymbol{\theta}(t)$ result in similar dynamical behavior, e.g., so that the process evolves on the same attractor, then the probability of recurrences will be higher, and so the similarity between rows or columns of the RP can be used as the basis for TVP reconstruction.

\citet{casdagliRecurrencePlotsRevisited1997} observed that, for appropriate TDE parameters, the RP of a TVP-driven non-stationary process approximates the RP of the TVP itself.
\citet{tanioReconstructionDrivingForces2009} later showed how to more reliably reconstruct a TVP from a RP by using combinatorial optimization to permute the RP so as to order the time-series points according to the similarity of their dynamics, thereby obtaining an approximate, monotonic transformation of the original TVP. 
\citet{hirataReproductionDistanceMatrices2008} developed another method for reconstructing time series from RPs by first creating a weighted graph based on neighborhood overlap, then computing all-pairs shortest paths using Dijkstra's algorithm, \cite{dijkstraNoteTwoProblems1959} and then applying multidimensional scaling \cite{meadReviewDevelopmentMultidimensional1992} to reconstitute the original time series.
This method also works when infinite-dimensional TDE is used for RP construction, and can be used for \problemname{} if the RP is coarse-grained prior to applying the reconstruction procedure.
\cite{hirataParsimoniousDescriptionPredicting2015,
hirataDimensionlessEmbeddingNonlinear2017}
Recently, \citet{gilpinRecurrencesRevealShared2023} proposed inferring a causal, driving process by first constructing a distance-based adjacency matrix representation, i.e., RP, of a multivariate time series and then computing the subleading eigenvector the the graph Laplacian of this matrix.

\subsection{Prediction error}
\label{sec:pred_error}

Consider the simple case of a non-stationary process $\mathcal{M}(\theta(t))$ where the effect of the scalar-valued TVP $\theta(t)$ is to translate the mean of the process.
For example, picture the Lorenz attractor (see Fig.~\ref{fig:Summary}a) translating up and down the $z$-axis with $\theta(t)$.
Suppose that a mean-sensitive predictive model is then trained using a portion of the associated time series, e.g., a single-step nearest-neighbor predictor (see Fig.~\ref{fig:Methods}d).
If the predictive model is then applied across time-series windows, one would expect the prediction error to vary according to changes in the mean of the process under the influence of $\theta(t)$.
Moreover, since the prediction error varies with $\Delta \theta$, one could use the time series of prediction errors to infer $\theta(t)$.
More generally, the methods we group here under the category `prediction error' involve training a predictive model using part or all of a measured time series and then inferring a TVP based on prediction error computed locally across time-series windows, typically by taking $\hat{\theta}_t$ to be the time series of prediction errors.

Many of these methods start with phase-space reconstruction (where the phase space is the space of all possible states of a dynamical system \cite{kantzNonlinearTimeSeries2003}).
The most common approach is to use a scalar-valued time series to construct a \textit{time-delay embedding} (TDE), i.e., a vector time series where each vector contains dimension $d$ entries from the original time series, each separated by a time delay of $\tau$. \cite{kantzNonlinearTimeSeries2003}
The appeal of TDE stems from Taken's theorem, \cite{takensDetectingStrangeAttractors1981, sauerEmbedology1991} which provides conditions under which there exists a diffeomorphism (a differentiable, invertible bijection) between the manifolds of the reconstructed and true dynamics, noting that a diffeomorphism preserves topologically invariant properties.
For the purpose of \problemname{}, with a phase-space reconstruction in hand, one can then fit prediction models and compute prediction errors with the goal of tracking TVPs.

\citet{schreiberDetectingAnalysingNonstationarity1997} proposed testing for the presence of non-stationarity using `cross-prediction error', which quantifies how well a nearest neighbors predictor for one time-series segment performs on another time-series segment, similar to our example above. 
\citet{schreiberClassificationTimeSeries1997} then used cross-prediction error to perform \problemname{}.
They partition a time series into windows and construct a dissimilarity matrix of cross-prediction errors between windows.
Clustering is then performed and a TVP is inferred based on computed distances from clusters over time.
\citet{verdesNonstationaryTimeSeriesAnalysis2001} provided a more general formulation of the prediction error approach.
Starting with a TDE $X$, they consider a predictive model $f$ such that $\boldsymbol{x}_{t+1} \approx f(\boldsymbol{x}_{t}, \theta_t)$ and show that $\theta_t$ can be expressed analytically in terms of prediction error, given: (i) a local linear approximation; and (ii) a smooth dependence of $f$ with $\theta$.
\citet{szeligaExtractingDrivingSignals2002} later proposed that $f$ and $\hat \theta_t$ be learned jointly using a multilayer perceptron neural network.
The network is trained to perform one-step time-series prediction using a mean squared error (MSE) loss function, and $\hat\theta_t$ is treated as a trainable vector where smoothness is enforced via an additional loss function of MSE across adjacent values $\hat\theta_t$ and $\hat\theta_{t+1}$.
In follow up papers, \citet{szeligaModellingNonstationaryDynamics2003} and \citet{verdesOverembeddingMethodModeling2006} showed how to use over-embedding and validation data to cope with noisy time series and to select hyperparameters during model training.
An advantage of the formulation of \citet{verdesNonstationaryTimeSeriesAnalysis2001} is that the form of $f$ is not specified, allowing $f$ to be constructed using a wide range of predictive modeling techniques.
Subsequent authors have used prediction models based on echo state networks \cite{gunturkunSequentialReconstructionDrivingforces2010} and support vector machines. \cite{grinblatNonstationaryRegressionSupport2015} 
However, an important limitation is that most of the methods in this category of methods yield only a \textit{scalar-valued} prediction error, so they can resolve a single TVP $\hat{\theta}_t$, but in order to resolve multiple TVPs $\hat{\boldsymbol{\theta}}_t$ some additional demixing technique is required, such as single-channel blind source separation. \cite{jangProbabilisticApproachSingle2002}

\subsection{Phase-space partitioning}
\label{sec:phase_space}

Instead of examining a process globally, phase-space partitioning methods analyze local regions of phase space, most commonly by partitioning phase space into hypercubes, as illustrated in Fig.~\ref{fig:Methods}e.
Phase space is either reconstructed, e.g., using the method of TDE (see Sec.~\ref{sec:pred_error}), or else a multivariate time series is taken to define a phase space, with each variable treated as a separate dimension.
The key intuition is that local analyses, e.g., in contiguous regions of phase space, may be more sensitive to subtle fluctuations in statistical properties that only occur in certain regions of phase space.
These methods output a multivariate time series to which some form of dimension reduction is applied, making it possible to resolve multiple TVPs underlying a single time series.
Note that some of the methods listed here compute prediction error locally, and so they are conceptually closely aligned with prediction error methods (Sec.~\ref{sec:pred_error}).
A key distinction is that, as noted, the methods in this section analyze multivariate time series of locally computed metrics, allowing for the inference of multiple TVPs, in contrast to the scalar-valued prediction errors used by methods from the previous section.
Additionally, the methods in this section represent a historical thread of \problemname{} research, mostly from the field of mechanical engineering, and forms a coherent body of work.

Chelidze \textit{et al.} developed a \problemname{} method called phase space warping (PSW)
\cite{chatterjeeOptimalTrackingParameter2002,
      chelidzeDynamicalSystemsApproach2002,
      chelidzeIdentifyingMultidimensionalDamage2004,
      chelidzeDynamicalSystemsApproach2005,
      chelidzePhaseSpaceWarping2006,
      chelidzeReconstructingSlowtimeDynamics2008,
      songSlowTimeChangesHuman2009,
      liGeometryinformedPhaseSpace2023}
that partitions phase space and computes prediction error locally in each region, yielding an error vector that evolves over time windows, followed by dimension reduction via smooth orthogonal decomposition (SOD).
Hyperparameter optimization is a challenge for this method, and is discussed at length in \citet{liGeometryinformedPhaseSpace2023}.
Note that SOD and SFA are equivalent, sharing the generalized eigenvalue problem formulation: $\dot{C}U = CU\Lambda$, where $C$ and $\dot{C}$ are the autocovariance matrices of a time series and its first temporal derivative, respectively, and $U$ and $\Lambda$ are eigenvectors and eigenvalues. \cite{chelidzeDynamicalSystemsApproach2005,
      berkesSlowFeatureAnalysis2005a}
However, there can be subtle differences in the algorithmic implementation of each method; for example, SFA incorporates a PCA whitening step during which near-zero eigenvalues can be optionally discarded according to some threshold.  \cite{konenNumericStabilitySFA2009}

Epureanu \textit{et al.} proposed the method of sensitivity vector fields (SVF) 
\cite{epureanuParameterReconstructionBased2006,
hashmiSensitivityResonanceAttractor2006,
yinExperimentalEnhancedNonlinear2007,
yinStructuralHealthMonitoring2006,
slobodaSensitivityVectorFields2013,
slobodaMaximizingSensitivityVector2014}
which first partitions phase space then computes a vector of local divergence between trajectories in each time window compared to a reference window, followed by dimension reduction via PCA.
Although the PSW and SVF methods are similar, key differences are that the former constructs local predictive models and applies SOD dimension reduction, whereas the latter compares trajectories directly and uses PCA dimension reduction.

\citet{nguyenNewInvariantMeasures2015} noted the computationally intensive nature of PSW and SVF, and so proposed a fast method called characteristic distance (CD) which is is based on the Birkhoff ergodic theorem. \cite{hawkinsErgodicDynamicsBasic2021} 
CD selects a set of random points in phase space and then computes vectors of mean distance from each point, yielding a time series of mean distance vectors to which SOD dimension reduction is then applied.
Sloboda \textit{et al.}
\cite{slobodaBoundaryTransformationRepresentation2021,
    slobodaBoundaryTransformationVectors2022,
    slobodaDamageAssessmentAttractor2022,
    slobodaRefinementsBoundaryTransformation2022}
observed that PSW and SVF require various modeling and hyperparameter choices, and in response proposed the boundary transformation vector (BTV) method as a way to directly measure local attractor deformation.
BTV infers parameter variation by computing 2D Poincar\'e sections for a number of planes through phase space and then computing vectors of boundary deformation metrics across time windows.
Sloboda et al. were also influenced by \citet{carrollAttractorComparisonsBased2015} who proposed a density-based method which partitions phase space into bins in order to compute an empirical probability density, after which feature vectors are produced through projection onto a polynomial basis, followed by TVP reconstruction using the Euclidean distance between the feature vector for each time window compared to a reference window.

Many of the phase-space partitioning \problemname{} methods have been used in applied research on fault detection and prognostication of the condition of industrial assets. \cite{chelidzeDynamicalSystemsApproach2002,
      gorjianjolfaeiPrognosticModellingIndustrial2022}
The idea is that for some non-stationary systems there exists a TVP that tracks a condition such as the gradual mechanical or electrical failure of a component.
In such cases, \problemname{} methods can be used to anticipate deteriorations in performance and eventual failure of the system.
Applications that have been examined include tracking the discharge of a battery,
\cite{chelidzeDynamicalSystemsApproach2002,
      chelidzeIdentifyingMultidimensionalDamage2004} 
crack growth in an oscillating system,
\cite{cusumanoDynamicalSystemsApproach2002,
      chelidzeDynamicalSystemsApproach2005,
      chelidzePhaseSpaceWarping2006,
      chelidzeReconstructingSlowtimeDynamics2008,
      liGeometryinformedPhaseSpace2023}
voltage input to a magneto-elastic oscillator,
\cite{nguyenNewInvariantMeasures2015}
behavior of an aeroelastic system,
\cite{yinStructuralHealthMonitoring2006}
the position of additive mass on a vibrating beam,
\cite{yinExperimentalEnhancedNonlinear2007}
variation in resistance in a chaotic Chua circuit,
\cite{slobodaSensitivityVectorFields2013,
      slobodaBoundaryTransformationRepresentation2021}
in addition to the biological case of tracking muscle fatigue in high-performance settings.
\cite{songSlowTimeChangesHuman2009,
      segalaNonlinearSmoothOrthogonal2011,
      bajelaniInfluenceCompressionGarments2022}

\subsection{Bayesian inference}
\label{sec:bayes}

When used to infer parameters, Bayesian inference starts with a prior probability distribution over possible parameter values (and/or models) and then computes a posterior distribution based on a probabilistic model of the conditional probability of the observed data given specific parameter values. \cite{gelmanBayesianDataAnalysis2013}
The posterior distribution can then be used to calculate the expected parameter value(s) and to quantify uncertainty in this estimate.
Bayesian methods have been used extensively to fit parameters to \textit{known} models, as in the case of the ODE inverse problem, and have been extended to handle TVPs.
\cite{sarkkaBayesianFilteringSmoothing2023,
gunturkunRecursiveHiddenInput2014,
arnoldApproachPeriodicTimevarying2018,
hauzenbergerFastFlexibleBayesian2022,
arnoldWhenArtificialParameter2023,
zhangParameterIdentificationFramework2023}
Bayesian methods have also been used for data-driven discovery, i.e., system identification, where the functional form of an ODE, PDE, or similar, is learned along with the corresponding static coefficients for each equation term.
\cite{northReviewDataDrivenDiscovery2023,
hirshSparsifyingPriorsBayesian2022a,
courseStateEstimationPhysical2023,
yuanMachineDiscoveryPartial2023}
A Bayesian approach to \problemname{} is illustrated in Fig.~\ref{fig:Methods}f.

Bayesian inference was used by \citet{smelyanskiyReconstructionStochasticNonlinear2005} to perform system identification for dynamical processes with an underlying SDE (or ODE) model.
Importantly, they developed an analytical, variational inference approach that converges quickly compared to Markov Chain Monte Carlo (MCMC) sampling.
The same research group subsequently showed that their technique can be applied to separate time-series windows in order to simultaneously resolve TVPs.
\cite{luchinskyInferentialFrameworkNonstationary2008,
      duggentoInferentialFrameworkNonstationary2008,
      luchinskyDynamicalInferenceHidden2008}
An important limitation of this method is that it assumes dense sampling and minimal measurement noise, which may not apply in practice.
Moreover, model selection is a problem, since the number of possible combinations of basis functions for the SDE model scales exponentially; for example, if a polynomial basis is used, there are $2^{d+1}$ possible combinations of polynomial terms up to order $d$ for a single variable.
Model selection was addressed through heuristic application of the Bayesian information criterion (BIC) and beam search.
\cite{morrisParameterStructureInference2005a} 
The restriction that the underlying model take the form of an SDE corresponds to only a subset of our \problemname{} formulation, but nonetheless includes a wide range of important processes.
Additionally, achieving simultaneous system identification \textit{and} \problemname{} is notable, and is an important direction for further research.

\section{Testing \problemname{} benchmark problems}
\label{experiment}

The relative performance of different \problemname{} algorithms is largely unknown, owing to a lack of comparative analyses across the diverse methodological literature.
In most studies, the performance of a given method is evaluated by comparing the TVP(s) $\hat{\boldsymbol{\theta}}_t$ inferred using a \problemname{} method against the ground truth $\boldsymbol{\theta}_t$ from a numerical simulation of a process.
The advantage of simulated data is that the ground truth is known, allowing performance of the TVP inference to be quantified using a metric such as Pearson correlation \cite{rodgersThirteenWaysLook} or normalized mean squared error (NMSE).\cite{verdesNonstationaryTimeSeriesAnalysis2001}
Many authors have focused the testing of their algorithms on simulations using well-characterized chaotic maps and flows, such as the Baker map, \cite{schreiberClassificationTimeSeries1997} the Tent map, \cite{guttlerReconstructionParameterSpaces2001, wiskottEstimatingDrivingForces2003, tanioReconstructionDrivingForces2009} the H\'enon map, \cite{guttlerReconstructionParameterSpaces2001, verdesOverembeddingMethodModeling2006} the Mackey--Glass system, \cite{guttlerReconstructionParameterSpaces2001, niknazarVolumetricBehaviorQuantification2017a} and the R\"{o}ssler system. \cite{hirataReproductionDistanceMatrices2008, nguyenNewInvariantMeasures2015}
Notably, the logistic map and the Lorenz process are the most commonly used processes for evaluating \problemname{} methods across the publications that we surveyed.
\cite{casdagliRecurrencePlotsRevisited1997,
      verdesNonstationaryTimeSeriesAnalysis2001,
      chatterjeeOptimalTrackingParameter2002,
      szeligaExtractingDrivingSignals2002,
      szeligaModellingNonstationaryDynamics2003,
      wiskottEstimatingDrivingForces2003,
      verdesOverembeddingMethodModeling2006,
      gunturkunSequentialReconstructionDrivingforces2010,
      slobodaSensitivityVectorFields2013,
      grinblatNonstationaryRegressionSupport2015,
      niknazarVolumetricBehaviorQuantification2017a,
      wangRobustnessDrivingForce2023} 
The common assumption is that these canonical, nonlinear, chaotic systems offer challenging cases that are useful for assessing the performance of \problemname{} algorithms.
However, since so many \problemname{} methods perform well on the same test problems, it raises suspicion about whether these are truly challenging problems on which strong performance is meaningful, or whether similar performance could be achieved by simpler baseline methods (which have not yet been adopted as standard practice within the literature).

To examine the suitability of the logistic map and Lorenz processes as \problemname{} benchmarking problems, we performed numerical experiments comparing the performance of several representative \problemname{} methods to a simple baseline method that estimates parameter variation as a time series of statistics, either mean or standard deviation, computed over sliding windows.
Undermining their use as meaningful test problems, we find that good \problemname{} performance in the logistic map and Lorenz process cases can be trivially achieved using our baseline method (Sec.~\ref{experiment1}).
Accordingly, we then introduce new, more challenging problems on which the same \problemname{} methods exhibit weaker performance, but for which we can still find statistical time-series features that perform well (Sec.~\ref{experiment2}).

\begin{figure*}
\begin{subfloatrow}
\includegraphics[width=0.5\textwidth]{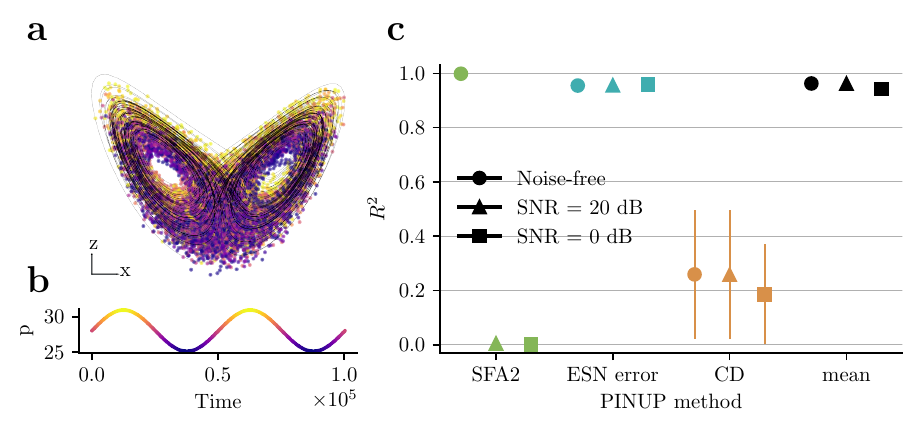}
\includegraphics[width=0.5\textwidth]{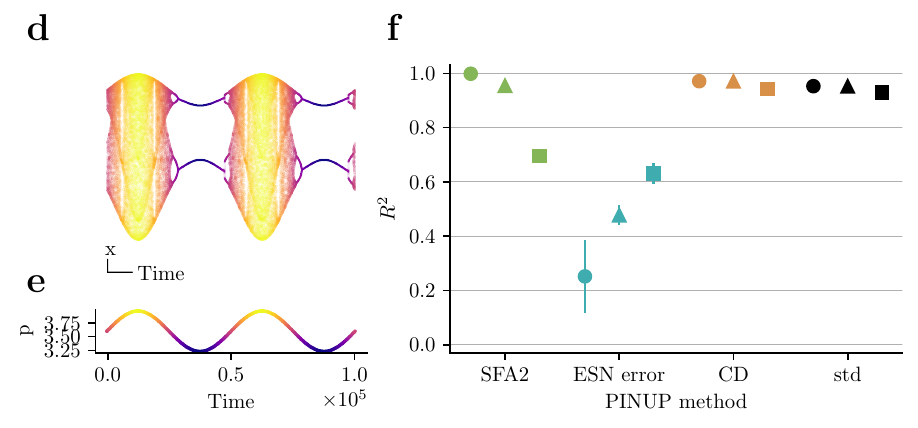}
\end{subfloatrow}

\begin{subfloatrow}
\includegraphics[width=0.5\textwidth]{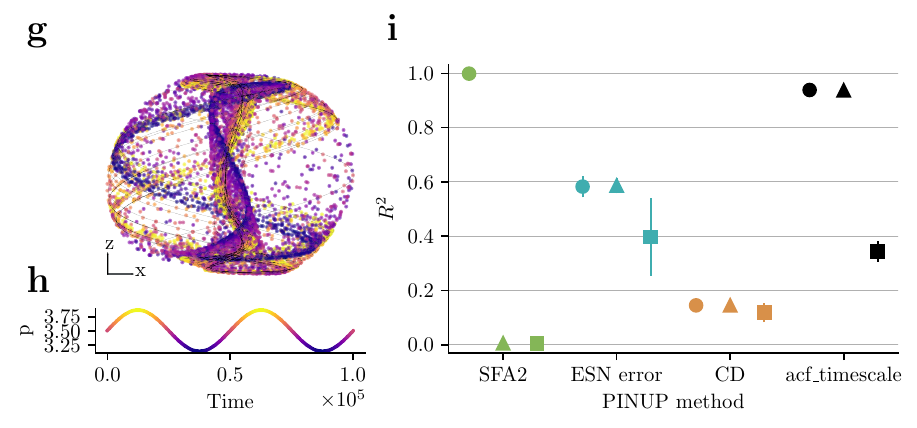}
\includegraphics[width=0.5\textwidth]{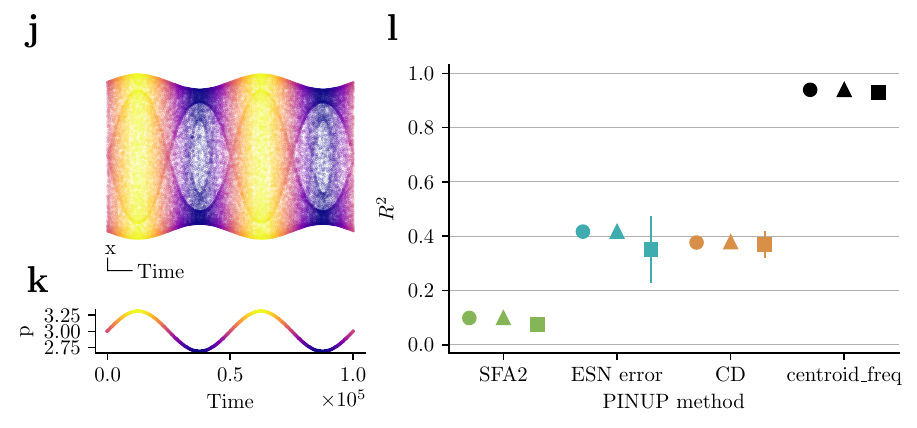}
\end{subfloatrow}[H]

\caption{\label{figExperiment}
\textbf{Individual time-series features match or exceed the TVP reconstruction accuracy of several \problemname{} methods across four chaotic systems.}
The processes (and varied parameters) are:
\textbf{a-c.} the Lorenz process (parameter $\rho$);
\textbf{d-f.} the logistic map (parameter $r$);
\textbf{g-i.} the Langford process (parameter $\omega$); and
\textbf{j-l.} the sine map (parameter $r$); each of which displays results from a parameter inference experiment.
A noise-free example of each process is visualized (in panels \textbf{a}, \textbf{d}, \textbf{g}, and \textbf{j}) and colored according to the corresponding TVP value at each point, noting that the flows are visualized as 2-dimensional spatial projections, whereas the 1-dimensional maps are visualized using space and time.
The time course of the associated sinusoidal TVP is visualized for each system (\textbf{b}, \textbf{e}, \textbf{h}, and \textbf{k}) with color corresponding to the TVP value for comparison to the corresponding process plots.
The numerical experiment result panels (\textbf{c}, \textbf{f}, \textbf{i}, and \textbf{l}) show mean and standard deviation of TVP reconstruction accuracy (Pearson $R^2$) for each method across multiple trials using two levels of signal-to-noise-ratio (SNR) ($0$ and $20$ dB) and a noise-free condition.
The \problemname{} methods were quadratic slow feature analysis (SFA2), a prediction-error-based method using echo state networks (ESN error), and characteristic distance (CD).
The single time-series features that were used were mean, standard deviation (\texttt{std}), the first time lag at which autocorrelation function falls below $1/e$ (feature name: \texttt{acf\_timescale} \cite{lubbaCatch22CAnonicalTimeseries2019a}), and the centroid of the power spectrum (feature name: \texttt{centroid\_freq} \cite{lubbaCatch22CAnonicalTimeseries2019a}).
}
\end{figure*}

\subsection{Testing non-stationary logistic map and Lorenz systems}
\label{experiment1}

We aimed to test whether strong \problemname{} performance on the non-stationary logistic map and Lorenz processes could be achieved by a simple, baseline method, relative to that of several existing \problemname{} methods.
To do this we first generated time series from each of the non-stationary logistic map and the Lorenz processes.
To demonstrate the comparative performance of several existing methods, \problemname{} was conducted using methods chosen from the dimension reduction, prediction error, and phase-space partitioning categories: quadratic SFA (SFA2, a dimension reduction method);\cite{wiskottEstimatingDrivingForces2003} smoothed mean prediction errors from a bank of echo state networks (ESN error, a prediction error method);\cite{gunturkunSequentialReconstructionDrivingforces2010} and characteristic distance (CD, a phase-space partitioning method).\cite{nguyenNewInvariantMeasures2015}
We chose these methods on the basis of ease of implementation, as well as to achieve a spread of methods across the categories described in Sec.~\ref{sec:existingMethods}.
A more detailed description of how we implemented each method is in Appendix~\ref{appendixMethods}.
In order to evaluate the performance of existing algorithms relative to a simple baseline method for tracking non-stationary variation, we used an approach belonging to the `time-series features' category, comprising single time-series features---mean and standard deviation---computed across sliding windows.
The resulting time series of statistical features is taken to be the estimate $\hat \theta_t$.
We were primarily interested in demonstrating the \textit{existence} of statistics capable of tracking TVPs, so in each case we selected the best-performing example, of mean or standard deviation applied to a single variable, for comparison with the other methods.
Since three of the methods---ESN error, CD, and our baseline---utilize time-series windows for the purpose of either smoothing or computing statistics, we adopted the same window length $10^3$ for each method, judging that this was sufficiently small relative to the slow TVPs to resolve parameter variation while being sufficiently large to confer robustness to noise.

We simulated each non-stationary process using a slow, sinusoidal TVP with period $T/2$, where $T$ was the total Lorenz process integration time or the number of logistic map iterations, respectively.
The amplitude of sinusoidal parameter variation was $\pm 10\%$ relative to the default value of the parameter that was varied for each system.
We note that qualitatively similar results are obtained for other functional forms for the slow parameter and across a range of values of low amplitude parameter variation.
We then evaluated the performance of each \problemname{} method by computing Pearson $R^2$ between the ground truth and inferred TVPs.
Mean $R^2$ was computed across 20 trials for each of three different levels of additive, Gaussian, observation noise: at signal-to-noise ratios (SNRs) of 0 and 20 dB, in addition to a noise-free condition.
In Appendix~\ref{appendixProcesses} we detail the equations, parameter settings, initial conditions, and functional form of the TVPs that were used for each simulated process.


The results of our experiments are shown in Figs~\ref{figExperiment}a-f, which for the Lorenz and logistic map systems visualize the process, the TVP, and the TVP reconstruction accuracy of each \problemname{} method.
Of the \problemname{} methods, ESN error performed well on the non-stationary Lorenz process (Fig.~\ref{figExperiment}c) and SFA2 and CD performed well on the non-stationary logistic map (Fig.~\ref{figExperiment}f).
SFA2 also achieved near-perfect performance on the Lorenz process in the noise-free condition but failed under both noise conditions.
In contrast, our simple baseline approach, which tracks either the mean or standard deviation across sliding windows, achieved high accuracy ($R^2 > 0.9$) across all noise conditions.

To understand why a simple windowed distributional feature can track the non-stationary variation underlying each of these nonlinear chaotic systems, first observe that when parameter values are mapped onto a 2-dimensional projection of the Lorenz phase space (Fig.~\ref{figExperiment}a) we see that the TVP induces a translation of the attractor along the $z$-axis.
Such translation in space is easily tracked by computing a mean statistic.
Similarly, it is visually evident that the variance of the logistic map is modulated by the corresponding sinusoidal TVP (Fig.~\ref{figExperiment}d).
The success of this simple baseline method undermines the apparent difficulty of these classical benchmark problems, which are ubiquitous in the \problemname{} literature and which are often (and incorrectly) used to evidence the usefulness of a given \problemname{} algorithm.
Even though chaotic processes like the Lorenz process and logistic map under the influence of a TVP may exhibit variation in nonlinear properties, our experiments show that using nonlinear statistics is not required for strong \problemname{} performance.
For these two systems this is because the parameter variation doesn't \textit{uniquely} affect nonlinear structure, but also varies simple properties---namely, the first two moments of the distribution.

\subsection{Finding challenging problems on which to evaluate \problemname{} methods}
\label{experiment2}

Given that the Lorenz process and logistic map \problemname{} problems can be trivially solved by tracking mean and variance, respectively, we next aimed to find more challenging alternative test problems that may be more suitable for demonstrating impressive \problemname{} performance.
Specifically, we aimed to demonstrate the existence of non-stationary processes for which accurate TVP reconstruction is possible using some \problemname{} methods, but for which the simple baseline method of windowed computation of mean or variance fails.
To do this we manually searched through the chaotic processes collated by \citet{sprottChaosTimeSeriesAnalysis2001} (comprising 29 maps and 33 flows) and \citet{gilpinChaosInterpretableBenchmark} (comprising 131 flows).
For each such system, we followed the comparative methodology of Sec.~\ref{experiment1}, i.e., varying a single parameter sinusoidally by $\pm 10\%$ of the default value, then quantifying the performance of our selected \problemname{} methods using Pearson $R^2$.
To discover whether there exist problems for which other time-series features beyond mean and standard deviation can track TVPs, we also performed the same sliding-window approach using 22 additional time-series features from the CAnonical Time-series CHaracteristics (\textit{catch22}) feature set. \cite{lubbaCatch22CAnonicalTimeseries2019a}

Here we present two processes, the Langford process and sine map, for which we found TVP inference to fail using sliding-window mean or standard variation, but for which at least one other method performs well (among SFA2, ESN error, CD, or one of the sliding-window \textit{catch22} features).
Our comparative results for these two processes are shown in Figs~\ref{figExperiment}g-l.
Weak performance on both the Langford and sine map processes was seen for all three of SFA2, ESN error, and CD (Figs~\ref{figExperiment}i,l), in addition to mean and standard variation (not shown).
The only exception to this was that SFA2 performed very well on the Langford process in the noise-free condition, similar to what was seen for the Lorenz process.
Observe that neither process exhibits obvious translation under the influence of the TVP (see Figs~\ref{figExperiment}g, j), consistent with the weak performance of sliding-window mean.
These are problems that require more algorithmic sophistication than simply computing mean and variance.
But these are not unsolvable problems---indeed, in each case we identified a single statistic from the \textit{catch22} feature set that could track the statistical changes induced by underlying parameter variation with reasonable accuracy (i.e., $R^2 > 0.8$) over two or more noise levels.
For the non-stationary Langford process, a statistical measure of timescale based on the decay of the autocorrelation function (\textit{catch22} feature name: \verb|acf_timescale|) successfully tracked parameter variation, albeit with a steep decline in performance with increasing noise, and for the non-stationary sine map a measure of the centroid of the power spectrum (\textit{catch22} feature name: \verb|centroid_freq|) performed well over all three noise levels.
A benefit of using simple time-series statistics is that they point towards an interpretable theory of what is changing in either process (i.e., autocorrelation and the power spectrum, respectively).

Taken together, these findings demonstrate the existence of more challenging \problemname{} problems, for which accurate TVP reconstruction is possible but is not achieved by the first two moments of the distribution or by a set of conventional methods.
If new \problemname{} methods can be developed that are capable of handling a range of such challenging problems, this will enable the inference of parameter variation for a wider range of non-stationary phenomena.

\section{Discussion}
\label{discussion}

Progress on the \problemname{} problem has been hindered by the fragmentation of the methodological literature across multiple disciplinary settings and journals, to the extent that many of the published \problemname{} methods reviewed here were developed with minimal knowledge of the progress made on the same problem in different fields.
We have made headway on this issue by providing the first overview of this interdisciplinary literature: formulating the \problemname{} problem class (in Sec.~\ref{sec:problemFormulation}) and organizing the literature into different categories of \problemname{} methods (in Sec.~\ref{sec:existingMethods}).
This shared terminology and set of reference algorithms will enable greater interaction and sharing of ideas between disciplines, while also fostering a systematic, comparative, methodological approach (across methods and problems) in future research.
In turn, we hope that such interdisciplinary, comparative work will enable progress on the \problemname{} problem.

A similar lack of systematic comparison has been faced by other algorithmic fields, such as computer vision, \cite{deng2009imagenet} time-series classification, \cite{keoghNeedTimeSeries2003} forecasting, \cite{makridakisAccuracyForecastingEmpirical1979, makridakisStatisticalMachineLearning2018} and clinical biomarker discovery, \cite{bronTenYearsImage2022, trautInsightsAutismImaging2022} and has been addressed in a range of ways.
For example, progress in time-series classification has been assisted by:
(1) developing libraries of time-series classification methods;
\cite{loningSktimeUnifiedInterface2019}
(2) curating repositories of benchmark problems with agreed performance metrics;
\cite{bagnallUEAMultivariateTime2018a}
and (3) conducting systematic comparisons of methods on these particular problems.
\cite{bagnallGreatTimeSeries2017,
    ruizGreatMultivariateTime2021}
We echo the `call to arms' issued by \citet{keoghNeedTimeSeries2003} that the culture of reviewers and scientists needs to move towards testing new methods on a broad set of challenging problems with comparison to a range of other methods, including simple, baseline approaches.

The \problemname{} literature has been in need of simple, baseline methods, to serve as a benchmark against which the performance of new methods can be compared.
The claimed usefulness of a new algorithm is justified by its superior performance relative to simpler baseline alternatives, and the field can only reasonably claim to be working on challenging problems when such simple methods fail.
We propose that mean or standard deviation computed across sliding time-series windows provides such a baseline.
Crucially, we found that this simple approach is comparable or even superior to several existing methods applied to the non-stationary Lorenz and logistic map processes (Sec.~\ref{experiment1}).
Indeed, for many of the non-stationary chaotic systems that we searched through in Sec.~\ref{experiment2}, either mean or standard deviation performed well.
A similar phenomenon has been observed in the time-series classification literature, where complicated algorithms sometimes fail to outperform classification using only mean and variance, even though these distributional properties are insensitive to the time-ordering of the data. \cite{hendersonNeverDullMoment2023}
Benchmarking on problems that can be solved using mean or variance has the downside of being unable to distinguish algorithms that track simple, distributional properties from those that can track more subtle forms of non-stationary variation.
By evaluating performance on new, more challenging problems, researchers in the \problemname{} field can more convincingly demonstrate advances in algorithmic performance.

Addressing this need, we performed a wide search across time-varying systems and demonstrated the existence of difficult problems for which TVP inference is possible, but where several existing methods (in addition to sliding-window mean and standard deviation) fail (Sec.~\ref{experiment2}).
In particular, we highlight two such problems: the non-stationary Langford and sine map processes.
Additional problems of this type could be found by applying our comparative methodology across a range of non-stationary processes.
More challenging \problemname{} problems such as these provide a valuable starting point for future research.
Harking back to the definitions that opened this paper, we speculate that an outstanding challenge for the \problemname{} field may be to achieve accurate TVP inference for processes that exhibit weak but not strong stationarity, i.e., where non-stationarity manifests through variation in joint probabilities but not through variation in mean or variance.

For each of these two, more challenging \problemname{} problems, we identified a single time-series statistic from the \textit{catch22} feature set that achieved accurate TVP reconstruction using the same sliding-window approach that we used for mean and variance (Sec.~\ref{experiment2}).
This simple sliding-window time-series feature-based approach is straightforward to implement, parsimonious, and enables the interpretation of non-stationary dynamics in terms of the variation of statistical properties over time.
The fact that this simple sliding-window approach can achieve strong performance where other methods fail presents a promising direction for future research; for example, one could refine feature selection or apply dimension reduction in the space of statistical features.
A downside of using statistical features is that a highly-sampled time series is needed to enable robust statistical estimates.
We expect the performance of \problemname{} methods to depend on a range of factors including sampling rate, the timescale of variation of the TVP relative to the observed process, the amplitude of parameter variation, and the contribution of dynamical and measurement noise. 
Future work could explore the strengths and weaknesses of different \problemname{} methods in the presence of variation in these factors.

By adopting a systematic approach and testing on appropriately difficult problems, \problemname{} researchers can both drive progress and help readers to discern genuine advances in the field.
Improved performance on tracking non-stationarity from time-series data may enable advances on a range of related time-series analysis problems for which conventional methods assume stationarity, including forecasting, \cite{wangNonstationaryTimeSeries2013, wangNonstationaryTimeSeries2015a} and system identification. \cite{luchinskyInferentialFrameworkNonstationary2008, northReviewDataDrivenDiscovery2023}
Taken together, our review and numerical results lay a foundation for interdisciplinary progress on the \problemname{} problem and the broader study of non-stationary phenomena.

\section*{Author Declarations}
\subsection*{Conflict of Interest}

The authors have no conflicts to disclose.


\section*{Data Availability Statement}

The data that support the findings of this study are openly available at \url{https://github.com/DynamicsAndNeuralSystems/pinup_paper}.

\newcommand{\singleappendix}[1]{%
  \appendix
  \section*{#1}
  \addcontentsline{toc}{section}{#1}
  \stepcounter{section}
}
\singleappendix{Appendix}
\label{appendix}

Here we detail the non-stationary processes and methods that were used for the numerical experiments described in Sec.~\ref{experiment} and shown in Fig.~\ref{figExperiment}.

\subsection{Non-stationary processes}
\label{appendixProcesses}

Non-stationary time series were generated, either through integration of a flow $\frac{d}{dt}x = f(x, \theta)$, given some initial state $x_0$, or iteration of a map $x_{t+1} = f(x_t, \theta_t)$.
In either case, the functional form of the TVP was a two-period sinusoid,
$$\theta(t) = p\left[1 + \alpha \sin\left(\frac{4\pi t}{T}\right)\right]\,,$$
where $p$ is the default parameter value, $\alpha$ reflects proportional variation, and $T$ was the total integration time or number of iterations, respectively. 
For all systems we used $\alpha = 0.1$, corresponding to sinusoidal variation of $\pm 10\%$ around the default parameter value $p$.
For ODE integration, the Runge--Kutta--Fehlberg (RKF45) method was used.
\cite{fehlbergLoworderClassicalRungeKutta1969}
The models that were used are described below.

\subsubsection{Iterative Maps}

For the logistic map 
\cite{maySimpleMathematicalModels1976} with equation:
\begin{align}
    x_{n+1} &= r x_n (1 - x_n)\,,
\end{align}
we used a parameter value of $r = 3.6$ and initial condition $x_0 = 0.6$. For the TVP we varied $r$.

For the sine map 
\cite{sprottChaosTimeSeriesAnalysis2001} with equation:
\begin{align}
    x_{n+1} &= r \sin(\pi x_n)\,,
\end{align}
we used a parameter value of $r = 3.0$ and the initial condition $x_0 = 0.6$. For the TVP we varied $r$.

For both systems, we iterated to simulate time series of length $10^5$ samples.

\subsubsection{Flows}

The Lorenz process is a model of atmospheric convection
\cite{lorenzDeterministicNonperiodicFlow1963} with equations:
\begin{align}
    \dot x &= \sigma(y - x)\,, \\
    \label{eq:lorenz}
    \dot y &= x(\rho - z) - y\,, \\
    \dot z &= xy - \beta z\,,
\end{align}
where parameters were set to values $\sigma=10$, $\rho=28$, and $\beta = 8/3$ and we used the initial condition $(x_0,y_0,z_0)= (-9.79, -15.04, 20.53)$ from Gilpin.
\cite{gilpinChaosInterpretableBenchmark} For the TVP we varied $\rho$.

The Langford process yields a torus-like attractor
\cite{langfordNumericalStudiesTorus1984} with equations:
\begin{align}
    \dot x &= (z - \beta)x - \omega y\,, \\
    \dot y &= \omega x + (z - \beta)y\,, \\
    \dot z &= \lambda + \alpha z - \frac{z^3}{3} - (x^2 + y^2)(1 + \rho z) + \varepsilon z x^3\,,
\end{align}
where parameters were set to values $\alpha = 0.95, \beta = 0.7, \lambda = 0.6, \omega = 3.5, \rho = 0.25,$ and $\varepsilon = 0.1$ and we used the initial condition $(x_0,y_0,z_0)= (-0.78, -0.63, -0.18)$ from Gilpin. \cite{gilpinChaosInterpretableBenchmark} For the TVP we varied $\omega$.

For both Lorenz and Langford systems, integration was performed over $1000$ time steps, sampling at intervals of $0.01$ time units.

\subsection{Methods}
\label{appendixMethods}

For each system, \problemname{} was performed using each of the below methods (SFA2, ESN, CD, and statistical time-series features) for two levels of additive Gaussian noise, yielding SNRs of 0 and 20 dB, in addition to a noise-free condition.
SNR was calculated as $10 \log_{10}(\sigma^2_s/\sigma^2_n)$, where $\sigma^2_s$ is the variance of the signal and $\sigma^2_n$ is the variance of the noise.

\subsubsection{Quadratic slow feature analysis (SFA2)}

This method is an example of a dimension reduction approach to \problemname{} (see Sec.~\ref{sec:dim_red}).
Following Wiskott, \cite{wiskottEstimatingDrivingForces2003} given a time series, we first applied time-delay embedding (TDE) with dimension $m$ and delay $\tau$, then applied a quadratic polynomial basis expansion, followed by single-component SFA, to obtain a parameter estimate $\hat \theta_t$. 
We set $\tau$ to the time delay corresponding to the first zero-crossing of the autocorrelation function (ACF), and in the setting of multivariate data selected the minimum such value computed across time-series variables.
To set $m$, using Wiskott's heuristic,\cite{wiskottEstimatingDrivingForces2003} we performed a search over integers $m \in \{1,\dots,20\}$ and selected the value of $m$ for which the mean squared value of the first temporal derivative of the TVP was minimized.

\subsubsection{Echo state network prediction error (ESN)}

This method is an example of a prediction-error approach to \problemname{} (see Sec.~\ref{sec:pred_error}).
Since no out-of-the-box algorithm was available, we implemented a method inspired by the approach of \citet{gunturkunSequentialReconstructionDrivingforces2010}: given a time series we train 50 echo state networks of internal dimension 30 and then obtain the mean prediction error across the 50 networks.
The mean prediction error time series was smoothed to obtain a TVP estimate.
Rather than using adaptive filtering, we performed smoothing using a Gaussian convolutional filter to obtain a TVP estimate.
We used $\sigma = 166$ for the Gaussian kernel standard deviation hyperparameter, yielding a filter window of approximately 1000 steps between $\pm 3 \sigma$, to maintain consistency with the window sizes used in the CD and statistical time-series feature methods.

\subsubsection{Characteristic distance (CD)}

This method is an example of a phase-space partitioning approach to \problemname{} (see Sec.~\ref{sec:phase_space}).
Following the general approach of Nguyen and Chelidze, \cite{nguyenNewInvariantMeasures2015} given a time series $x_t$ of length $10^5$, 50 random points were uniformly sampled from a phase space hypercube bounding the points in $x_t$ but with boundaries extended by $\pm 10\%$ in each dimension to allow for sampling of points `outside' the distribution
A vector of Euclidean distances from each of the 50 random points was computed for each of the points in $x_t$.
A vector of 50 mean distances was then computed for contiguous, non-overlapping windows of length $10^3$.
Single-component SFA was then applied to obtain a TVP estimate.

\subsubsection{Statistical time-series features}

This simple baseline method is an example of a statistical time-series feature approach to \problemname{} (see Sec.~\ref{sec:stat_ts_features}), noting that it is not yet standard practice within the \problemname{} literature to use this method as a baseline comparison.
For each variable of each process, the time-series features from the \textit{catch22} library \cite{lubbaCatch22CAnonicalTimeseries2019a} were computed over sliding windows.
Specifically, each time series variable of length $10^5$ was partitioned into contiguous, non-overlapping windows of length $10^3$ and the features were computed for each window yielding feature time series of length $100$.


\bibliography{pinup}

\end{document}